\documentclass[acmsmall,authorversion]{acmart}
\AtBeginDocument{%
  }

\setcopyright{acmlicensed}
\copyrightyear{2025}
\acmYear{2025}
\acmDOI{}
\acmVolume{}
\acmNumber{}
\acmMonth{1}
\received{2024-12-13}
\received[Revised]{2025-01-12}

\acmConference{The GROUP '25 Workshop on ``Beyond Video-Conferencing'': Telepresence Technologies for Extending Expertise Reach and Specialised Skill Sharing}{January 12,
  2025}{Hilton Head Island, South Carolina}

\begin{document}

\title[Gesture-Aware Augmented Reality Video for Data-Rich Remote Presentations]{Video-Conferencing Beyond Screen-Sharing and Thumbnail Webcam Videos: Gesture-Aware Augmented Reality Video for Data-Rich Remote Presentations}

\author{Matthew Brehmer}
\email{mbrehmer@uwaterloo.ca}
\orcid{1234-5678-9012}
\affiliation{%
  \institution{University of Waterloo}
  \city{Waterloo}
  \state{Ontario}
  \country{Canada}
}

\renewcommand{\shortauthors}{Brehmer}

\newcommand{\cf}{cf.\ }

\newcommand{\bstart}[1]{\vspace{1mm} \noindent{\textbf{#1.}}}
\newcommand{\istart}[1]{\vspace{1mm} \noindent{\textit{#1.}}}
\newcommand{\bcstart}[1]{\vspace{1mm} \noindent{\textbf{#1:}}}
\newcommand{\bdef}[1]{\vspace{1mm} \noindent{\textbf{#1}}}
\newcommand{\idef}[1]{\vspace{1mm} \noindent{\texttt{#1}}}

\begin{abstract}
\textbf{Abstract}: Synchronous data-rich conversations are commonplace within enterprise organizations, taking place at varying degrees of formality between stakeholders at different levels of data literacy. In these conversations, representations of data are used to analyze past decisions, inform future course of action, as well as persuade customers, investors, and executives. However, it is difficult to conduct these conversations between remote stakeholders due to poor support for presenting data when video-conferencing, resulting in disappointing audience experiences. In this position statement, I reflect on our recent work incorporating multimodal interaction and augmented reality video, suggesting that video-conferencing does not need to be limited to screen-sharing and relegating a speaker's video to a separate thumbnail view. I also comment on future research directions and collaboration opportunities.
\end{abstract}



\begin{teaserfigure}
    \centering
    \includegraphics[width=1\textwidth]{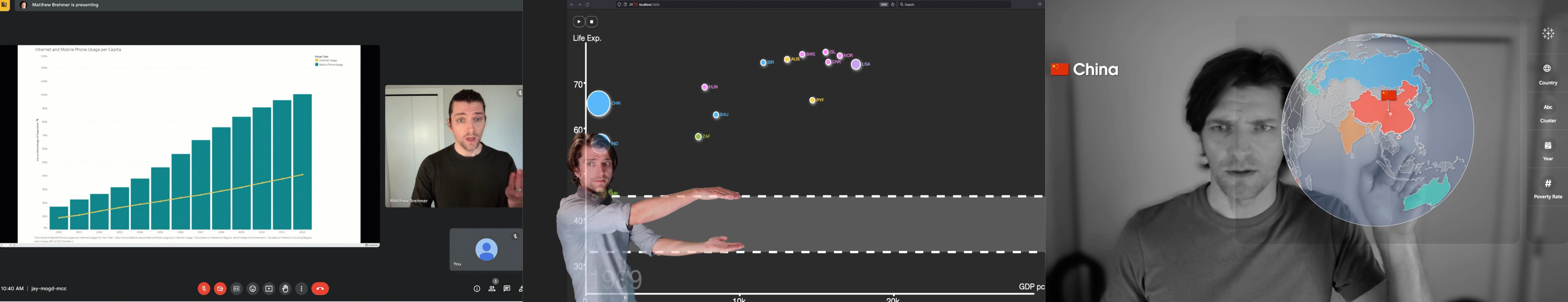}
    \caption{
        Three approaches to data-rich presentations for remote audiences. \textbf{Left}: the status quo of presenting data to remote audiences via screen-sharing~\cite{brehmer2021jam}. \textbf{Middle}: A frame from a webcam video in which a pose recognition model is used to control the state of a dynamic information display, green-screen composited behind the presenter~\cite{infohands2021}. \textbf{Right}: A frame from a webcam video in which continuous hand-tracking is used to control dynamic representations of information composited in the foreground of the video~\cite{hall2022augmented,tableauGestures}.
    }
    \label{fig:presenter_vs_audience}
\end{teaserfigure}


\maketitle

\section*{Position Statement}

\bstart{Personal Introduction \& Context} 
In 2019, I joined \href{https://www.tableau.com/}{Tableau} as a research scientist, where I was dedicated to designing and evaluating new experiences that would further the company's mission to \textit{help people see and understand data}.
My area of focus was to improve the ways by which people in enterprise organizations give presentations about data to their stakeholders, such as when analysts present the results of sales and marketing campaigns to product leadership, or when those in communication roles present earnings results to investors and executives. 
The COVID-19 pandemic narrowed this focus to the challenges associated with presenting data to remote audiences.
We learned in an interview study with 18 enterprise knowledge workers~\cite{brehmer2021jam} that neither slide presentation tools nor interactive business intelligence dashboard applications were appropriate for this activity; the former typically implied a tedious effort to create static presentation assets, precluding informal discussions about data, while the latter often gave audiences the impression that they were watching a software demonstration. 
With either approach, audiences would often be disengaged or uncertain as to what details to attend to; moreover, both approaches implied screen-sharing content with a spatially-separated thumbnail speaker video (e.g.,~\autoref{fig:presenter_vs_audience}-Left).


\bstart{Augmented Video Presentations}
Inspired in part by how broadcast journalists' use of dynamic displays of information to present data, we considered whether audiences would benefit from seeing a presenter co-located with their visual aids, in which they might use deictic body language with spatially-adjacent visual cues to direct their audiences' attention. 
However, most employees of enterprise organizations do not have a large touchscreen display or a visual effects team at their disposal, particularly when working remotely. 
What many enterprise employees do have is a webcam, so we investigated augmented video presentation techniques that composite webcam video with dynamic information displays, interactively controlled with pose recognition and hand-tracking provided by computer vision models (e.g., \href{https://teachablemachine.withgoogle.com/}{Teachable Machine}, \href{https://ai.google.dev/edge/mediapipe/solutions/guide}{MediaPipe}), and shared with remote audiences via a virtual camera application (e.g., \href{https://obsproject.com/}{Open Broadcaster Software / OBS}). 

Our first approach~\cite{infohands2021} involved green-screen compositing a dynamic information display behind the presenter (\autoref{fig:presenter_vs_audience}-Center), using a pose recognition model to update the display.
However, this required careful lighting and choreography, making for a rigid and unnatural presentation experience, in which the presenter would often appear distracted by repeatedly glancing at a reference monitor.

Our next approach proved to be more successful~\cite{hall2022augmented}, in which we composited semi-transparent charts in the video foreground and used continuous bimanual hand-tracking to perform both operational and expressive gestures that simultaneously modify the visual aids and direct audience attention; by mirroring the video, it became easy for presenters to coordinate their gestures relative to the displays.
This work would evolve into an application called \textit{Tableau Gestures} (\autoref{fig:presenter_vs_audience}-Right), which we demonstrated to customers at the 2023 and 2024 Tableau Conferences~\cite{tableauGestures}.

\bstart{Additional Modalities \& Interactive Authoring Support}
The initial gestural vocabulary of Tableau Gestures was restricted to revealing, comparing, and annotating data elements in the presenter's video foreground.
We realized that some of the transformations that presenters make to representations of data have no obvious gestural command, such as sorting, aggregating, or changing the color associations with categories of values. 
To address this gap, we added speech recognition to a variant of Tableau Gestures that would update the representations of data in the foreground in response to the utterance of template keyword phrases~\cite{MERCADOsrinivasan}; in doing so, a new challenge surfaced: the presenter must incorporate these utterances naturally and coherently into their spoken monologue. 

Most recently, another project resulted in another application dedicated to gesture-aware augmented video presentations about data.
Our open-source \textit{VisConductor} project~\cite{femi2024visconductor} considered a different gestural vocabulary, one dedicated to illustrating visuals that gradually animate over time, incorporating commands for foreshadowing the reveal of future data patterns and modulating the affective properties of animations.
Notably, it also incorporated a widget-based graphical interface for specifying both the placement of visual aids and gestural activation zones, as well as presenter interface that provided visual feedback during presentation delivery. 

\bstart{Community Development \& Concurrent Related Work}
In 2023, myself and a team of international collaborators hosted the first workshop on \textit{Multimodal Experiences for Remote Communication Around Data Online (MERCADO)} at the IEEE VIS Conference~\cite{mercado2023}, gathering those interested in the intersection of data visualization, remote / hybrid communication and collaboration, and multimodal input and output, including others interested in the prospect of augmented reality video for data-rich presentations for engaging remote audiences~\cite{MERCADOkristanto}.
This theme continued with our 2024 NII Shonan seminar on \textit{``Augmented Multimodal Interaction for Synchronous Presentation, Collaboration, and Education with Remote Audiences''}~\cite{shonanreport}, which broadened participation to those more closely aligned with SIGCHI communities (e.g., CHI, CSCW, ISS, UIST). 
Discussion at both events suggested a collective interest in approaches to augment or complement video-conferencing scenarios, as well as an interest in delineating scenarios that would be better served by emerging extended reality technologies, including those used in the field of \textit{immersive analytics}~\cite{saffo2023unraveling}.
I intend to continue to identify interests common to these intersecting groups at future events.

\bstart{Challenges \& Research Opportunities}
There are several open challenges with respect to augmenting existing video-conferencing experiences, and each can be extrapolated to general challenges in data-rich remote communication and collaboration~\cite{shonanreport}, which in turn echo some of the open challenges in collaborative immersive analytics~\cite{ens2021grand}.
The first is a challenge relating to the intersection of data and display technology: augmenting webcam video by compositing visuals may be appropriate for abstract two-dimensional charts, but this approach may not be viable when discussing data that has an inherent three-dimensional structure, requiring a representation that communicates volume and depth.
A second challenge pertains to scalability and the diversity of roles: the approaches described above may be appropriate for largely unidirectional communication between a presenter and an audience, but they may be less appropriate for scenarios of negotiation or group consensus building, in which multiple participants seek to interact with shared representations of data.
A third challenge is whether and how to leverage AI-based assistance in presentation scenarios; we have seen recent work incorporating AI-based content generation in online meetings~\cite{liu2023visual,xia2023crosstalk}, and extending these approaches to generate and manipulate shared visual representations of data in the presence of a diversity of roles and display types is an exciting direction for future research. 
Finally, evaluating these experiences will be challenging, particularly in the scenarios involve role asymmetry or device asymmetry, particularly if a subset of presenters and viewers participate via head-mounted displays instead of webcams.

\bstart{Beyond Video-Conferencing Does Not Mean Beyond Webcams} 
Many of the frustrations that people have with video-conferencing are arguably frustrations with the conventions of video-conferencing applications: the reliance on screen-sharing and a relegation of participants to thumbnail videos.
Some of this frustration can also be attributed to how people and organizations use video-conferencing applications: when webcams capture only a participant's face, or when many participants opt to participate passively without sharing their video. 
These frustrations are particularly apparent in information-dense presentations and meetings, where neither presenter or audience can have a satisfying experience. 
As excited as I am about new technologies (e.g., head-mounted augmented reality displays, robotics), most employees of enterprise organizations do not have access to (or the desire to use) these technologies, but they \textit{do} have webcams and microphones.
Using these affordable and unobtrusive devices, we can incorporate gesture and speech interaction in our applications as described above, as well as room and object recognition~\cite{Liao2022RealityTalkRS}.  
By experimenting with commodity hardware in these ways, we can extend their capabilities beyond how they are used in commercial video-conferencing applications.


\bibliographystyle{ACM-Reference-Format}
\bibliography{sample-base}

\end{document}